\documentclass{article}
\usepackage{spconf,amsmath,amssymb,graphicx,multirow,subcaption}
\usepackage{tikzpagenodes}

\newcommand{\atoms}{M}
\newcommand{\authorspace}{\hspace{1.0em}}
\newcommand{\channels}{C}
\newcommand{\dict}{D}
\newcommand{\dwell}{t}
\newcommand{\gt}{\mathbf{Y}}
\newcommand{\height}{H}
\newcommand{\loss}{\mathcal{L}}
\newcommand{\N}{\mathbb{N}}
\newcommand{\numel}{N}
\newcommand{\pnll}{\mathcal{P}}

\newcommand{\R}{\mathbb{R}}
\newcommand{\rate}{\mathbf{\Lambda}}
\newcommand{\rep}{A}
\newcommand{\rgb}{\mathbf{I}}
\newcommand{\rgbw}{\Omega}
\newcommand{\tv}{\text{TV}}
\newcommand{\width}{W}
\newcommand{\xrf}{\mathbf{X}}

\DeclareMathOperator*{\argmin}{arg\,min}
\newcommand{\etal}{\textit{et al.}~}

\title{Denoising Fast X-Ray Fluorescence Raster Scans of Paintings}
\name{H. Chopp$^{1*}$ \authorspace A. McGeachy$^{1}$ \authorspace M. Alfeld$^{2}$ \authorspace O. Cossairt$^{1}$ \authorspace M. Walton$^{1}$ \authorspace A. K. Katsaggelos$^{1}$ \thanks{We thank funding from NSF PIRE grant \#1743748: Computationally-Based Imaging
of Structure in Materials (CuBISM) for supporting our work.}}
 \address{$^{1}$ Northwestern University, Evanston IL, United States \\
      $^{2}$ Delft University of Technology, Delft, The Netherlands \\
      \small *HenryChopp2017@u.northwestern.edu}
\begin{document}
%
\maketitle

\begin{tikzpicture}[remember picture,overlay]
    \node[align=left] at ([yshift=-2.5em]current page text area.south) {\footnotesize © 2022 IEEE. Personal use of this material is permitted. Permission from IEEE must be obtained for all other uses, in any current or future media, including \\ \footnotesize reprinting/republishing this material for advertising or promotional purposes, creating new collective works, for resale or redistribution to servers or lists, or \\ \footnotesize reuse of any copyrighted component of this work in other works.};
  \end{tikzpicture}%
\begin{abstract}
Macro x-ray fluorescence (XRF) imaging of cultural heritage objects, while a popular non-invasive technique for providing elemental distribution maps, is a slow acquisition process in acquiring high signal-to-noise ratio XRF volumes.
Typically on the order of tenths of a second per pixel, a raster scanning probe counts the number of photons at different energies emitted by the object under x-ray illumination.
In an effort to reduce the scan times without sacrificing elemental map and XRF volume quality, we propose using dictionary learning with a Poisson noise model as well as a color image-based prior to restore noisy, rapidly acquired XRF data.
\end{abstract}
\section{Introduction}
\label{sec:intro}

In the growing field of applying scientific methods to cultural heritage research, macro x-ray fluorescence (XRF) imaging is frequently used as a non-invasive tool to analyze works of art.
This approach leverages the insights gained from XRF point analysis in providing elemental information on a per-pixel basis.
These elemental distribution maps provide information as to what chemical elements compose the layers of paint.
With these maps for example, an art conservator can better preserve paintings~\cite{janssens2016rembrandt}, or an art historian can deduce an artist's painting techniques---sometimes revealing hidden paintings~\cite{pouyet2020new}.

In macro XRF imaging, a source excites a small target area of the painting by irradiating it with x-rays.
An inner orbital electron can be ejected if the impinging x-ray has greater energy than the electron's binding energy.
An electron at an outer orbital then drops to fill the inner orbital vacancy by emitting a photon of energy equal to the energy difference of the orbitals.
Each element has characteristic orbital energy levels (and therefore a characteristic XRF spectrum).
A detector and digital post processor records and bins each photon according to its energy.

\begin{figure}[t!]
    \centering
    \includegraphics[width = 0.65 \linewidth]{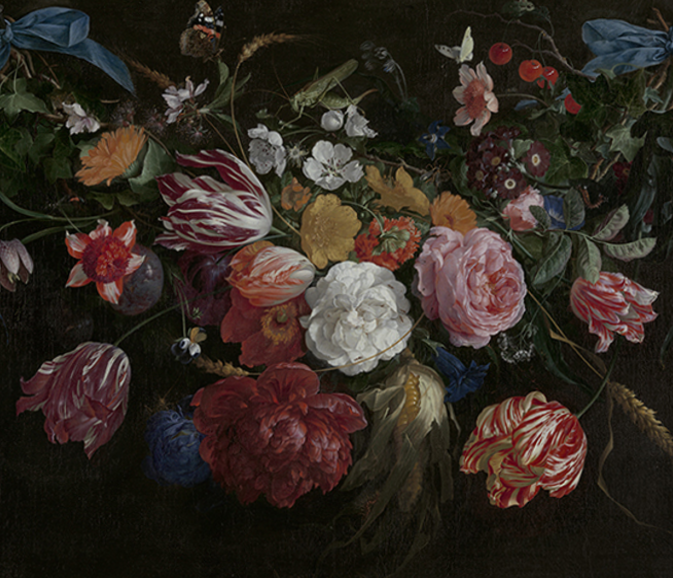}
    \caption{Jan Davidsz. de Heem's \textit{Bloemen en Insecten}, $49 \times 67$ cm, Royal Museum of Fine Arts Antwerp, inv. no. 54, oil on canvas.}
    \label{fig:rgb}
\end{figure}

While macro XRF is a powerful, increasingly popular technique, acquiring elemental maps for entire paintings with good signal-to-noise ratios, often translates to long acquisition times.
Depending on the painting size, spot size, and dwell time, it can take many hours to acquire the XRF volume.
Take as an example a modest painting of size $600 x 720$ mm.
If we specify a scan with spot size 1 mm\textsuperscript{2} and dwell time $0.2$ s/px, it would take exactly $1$ day to scan.
There are two problems in that (1) access to paintings often occur in short time windows when they are off-view, en route to other sites, etc., and (2) the x-ray exposure time should be minimized to best preserve the painting.

Analysis of XRF volumes uses photon count \textit{rates} instead of photon counts, as the dwell time can vary by scan.
These volumes are then separated into elemental maps using a least squares fit where the feature matrix is composed of known elemental XRF responses.
Before collecting XRF data, a trade-off between image quality (e.g. mean-square error (MSE)) and time must be taken into account: the longer the dwell time, the more accurate the measured photon count rates from the count-limited photon data.
Our goal here is to develop an XRF denoising algorithm where we test it on simulated scans at different dwell times based on real XRF data.
We focus on Jan Davidsz. de Heem's \textit{Bloemen en Insecten} as shown in figure~(\ref{fig:rgb}), the data of which has been generously shared by de Keyser \etal~\cite{dekeyser2017jan}.

\section{Related Work}
\label{sec:related}

Dictionary learning approaches frequently appear in XRF literature since each element emits a characteristic set of discrete fluorescent lines.
Limiting the number of spectral representations to the number of elements makes intuitive sense, as each pixel is then a linear combination of different elemental spectra.
Martins \etal proposed denoising XRF volumes using multivariate curve resolution-alternating least squares (MCR-ALS), a simple dictionary learning approach in the spectral domain to separate elemental compositions~\cite{martins2016jackson, martins2016piet}.
Kogou \etal used an unsupervised learning method called self-organizing maps (SOMs) that also extracts a set of spectral dictionary atoms to decompose the XRF volumes into a representative basis~\cite{kogou2021new}.
This method effectively uses k-means clustering to generate the set of dictionary endmembers.
More elaborate dictionary methods have been explored by Dai \etal whereby joint RGB and XRF dictionaries inpaint a spatially selective subsampled XRF volume~\cite{dai2020adaptive}.

Even though photons arrive according to a Poisson process~\cite{holmes1991acceleration}, each of these methods (implicitly) uses a white Gaussian noise model since the dwell times are assumed to be long.
This noise model was shown to be a good approximation in XRF denoising due to the central limit theorem, but can break down with short dwell times when white Gaussian noise is no longer an accurate approximation as our experiments show.

PURE-LET from Luisier \etal is an algorithm specifically for Poisson image denoising that minimizes the Poisson unbiased risk estimate in the Haar-wavelet domain~\cite{luisier2010fast}.
This method was originally published using tests on conventional images, MRI brain data, and fluorescence-microscopy of biological samples.
To the best of our knowledge, it has not been applied to XRF data, but is another tool that can be used as it partially addresses the concerns of current dictionary learning approaches for XRF denoising.

Our method merges the best characteristics of the two solution approaches: a spectral dictionary learning approach with a Poisson model (instead of a Gaussian model) to denoise XRF volumes.
An RGB image prior and sparsity coding are also used for denoising the data.

\section{Algorithm}
\label{sec:algo}

Assume for now that we have the XRF volume from a fast raster scan, $\xrf \in \N^{\height \times \width \times \channels}$ of height $\height$, width $\width$, and channels (\textit{i.e.} energy bins) $\channels$.
Each pixel has an identical dwell time $\dwell$.
The photons arrive with unknown underlying photon arrival rate $\rate \in \R^{\height \times \width \times \channels}_+$.
Additionally, assume we have an RGB image of the painting $\rgb \in \left[ 0, 1 \right]^{\height \times \width \times 3}$ registered with the XRF data.
We estimate the count rate $\rate^* \approx \rate$ using $\xrf$, $\rgb$, and $\dwell$ in our optimization formula detailed here.

\subsection{Formulation}

\begin{figure*}[ht!]
    \centering
    \newcommand{\imlen}{0.137}
    \setlength\tabcolsep{0.5pt}
    \renewcommand{\arraystretch}{0.25}
    \footnotesize
    \begin{tabular}{cccccccccccccc}
        \multicolumn{2}{c}{\small (a) \; Pb L3} &
        \multicolumn{2}{c}{\small (b) \; Cu K} &
        \multicolumn{2}{c}{\small (c) \; Ca K} &
        \multicolumn{2}{c}{\small (d) \; Co K} &
        \multicolumn{2}{c}{\small (e) \; As K} &
        \multicolumn{2}{c}{\small (f) \; Cl K} & 
        \multicolumn{2}{c}{\small (g) \; Si K} \\
        
        \multicolumn{2}{c}{\rotatebox{90}{\small Fast Scan} \includegraphics[width = \imlen\textwidth]{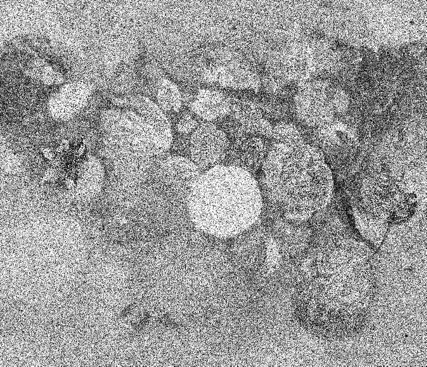}} &
        \multicolumn{2}{c}{\includegraphics[width = \imlen\textwidth]{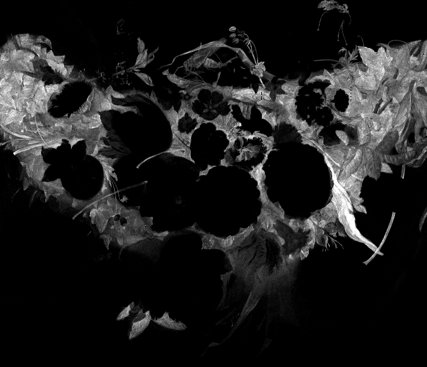}} &
        \multicolumn{2}{c}{\includegraphics[width = \imlen\textwidth]{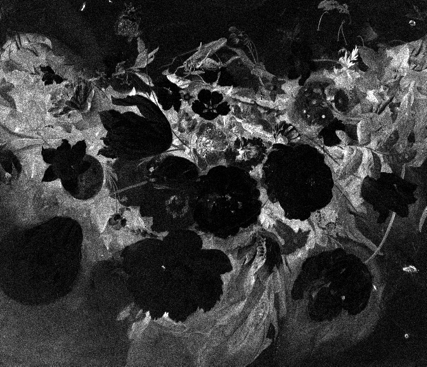}} &
        \multicolumn{2}{c}{\includegraphics[width = \imlen\textwidth]{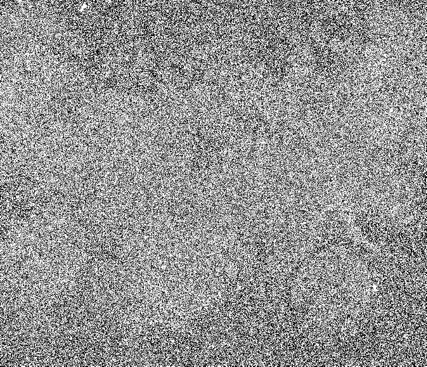}} &
        \multicolumn{2}{c}{\includegraphics[width = \imlen\textwidth]{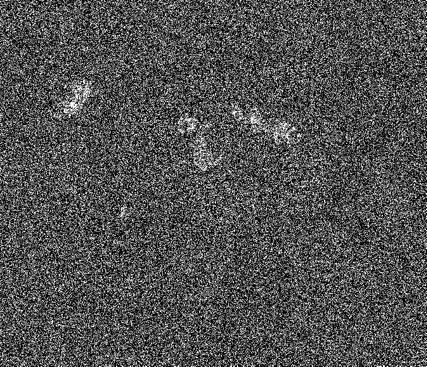}} &
        \multicolumn{2}{c}{\includegraphics[width = \imlen\textwidth]{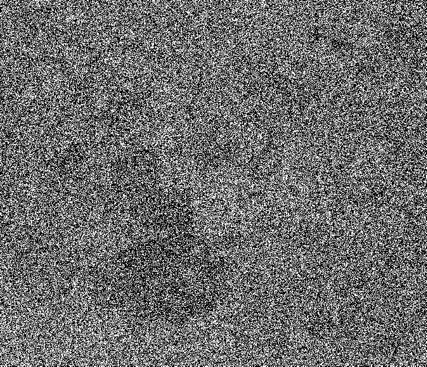}} &
        \multicolumn{2}{c}{\includegraphics[width = \imlen\textwidth]{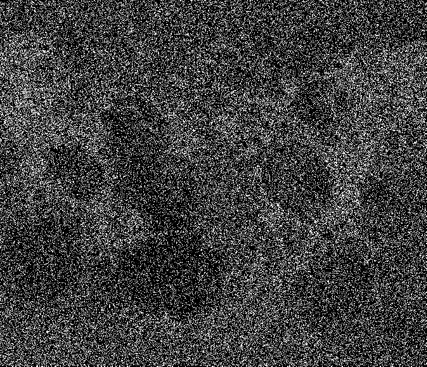}} \\
        
        \quad\; 898149 & (-10797) & \; 18724 & (-7328) & \quad 6208 & (-1245) & \quad 4116 & (-151.95) & \quad 336360 & (173.01) & \quad 312.40 & (-4.78) & \quad 137.24 & (0.7262) \\ \\
        
        \multicolumn{2}{c}{\rotatebox{90}{\small PURE-LET} \includegraphics[width = \imlen\textwidth]{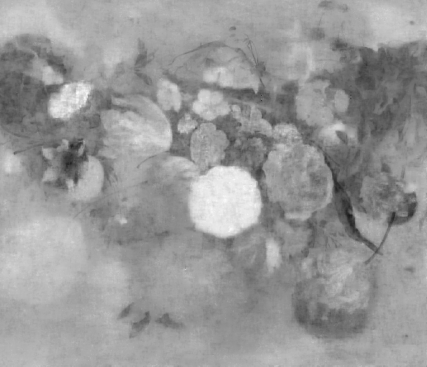}} &
        \multicolumn{2}{c}{\includegraphics[width = \imlen\textwidth]{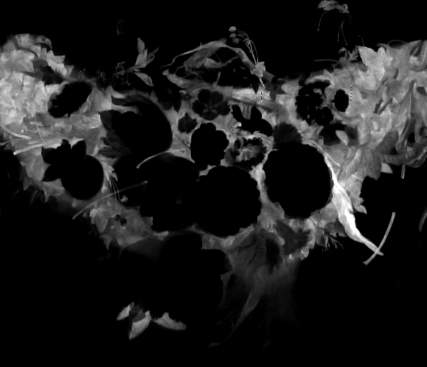}} &
        \multicolumn{2}{c}{\includegraphics[width = \imlen\textwidth]{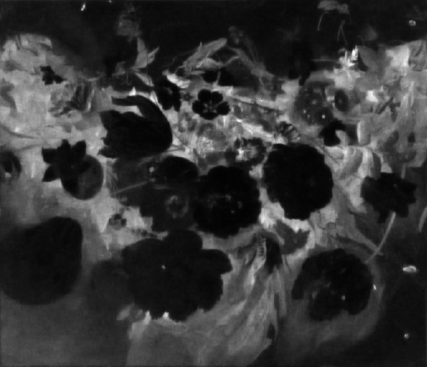}} &
        \multicolumn{2}{c}{\includegraphics[width = \imlen\textwidth]{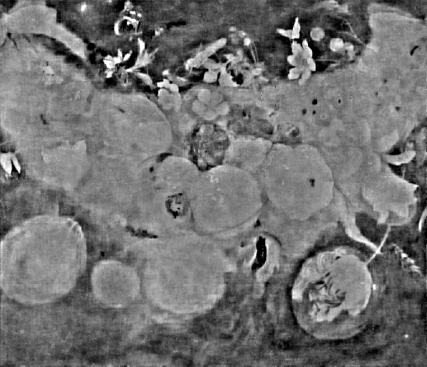}} &
        \multicolumn{2}{c}{\includegraphics[width = \imlen\textwidth]{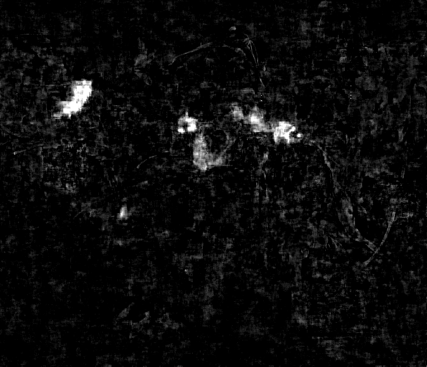}} &
        \multicolumn{2}{c}{\includegraphics[width = \imlen\textwidth]{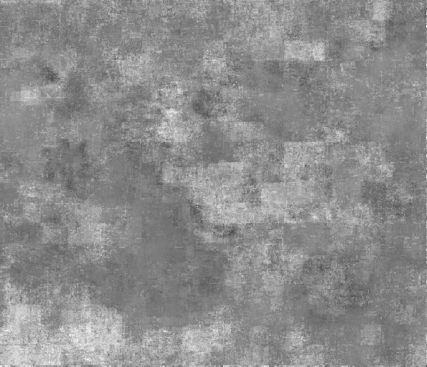}} &
        \multicolumn{2}{c}{\includegraphics[width = \imlen\textwidth]{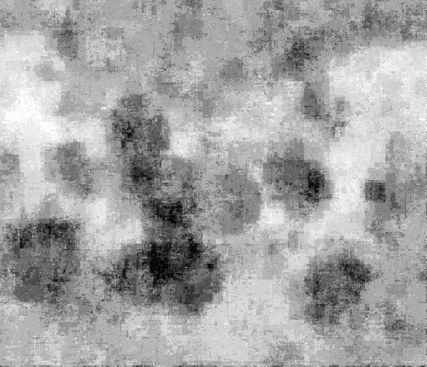}} \\
        
        \quad\; 36777 & (-11369) & \; 55864 & (-7326) & \quad \textbf{4912} & (\textbf{-1255}) & \quad 299 & (-183.82) & \quad \textbf{9108} & (\textbf{-103.04}) & \quad \textbf{22.55} & (\textbf{-14.99}) & \quad 12.89 & (-3.6336) \\ \\
        
        \multicolumn{2}{c}{\rotatebox{90}{\small MCR-ALS} \includegraphics[width = \imlen\textwidth]{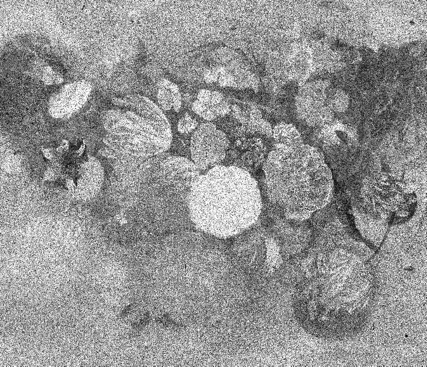}} &
        \multicolumn{2}{c}{\includegraphics[width = \imlen\textwidth]{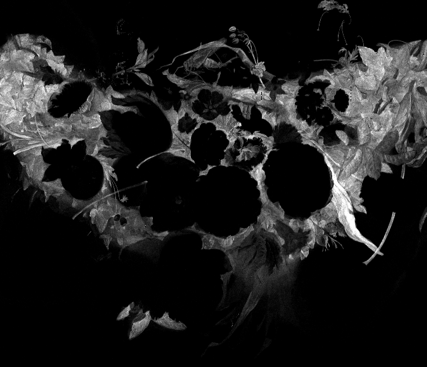}} &
        \multicolumn{2}{c}{\includegraphics[width = \imlen\textwidth]{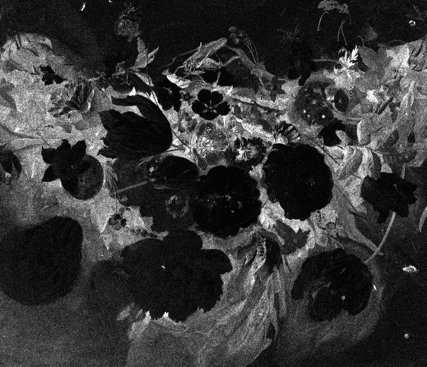}} &
        \multicolumn{2}{c}{\includegraphics[width = \imlen\textwidth]{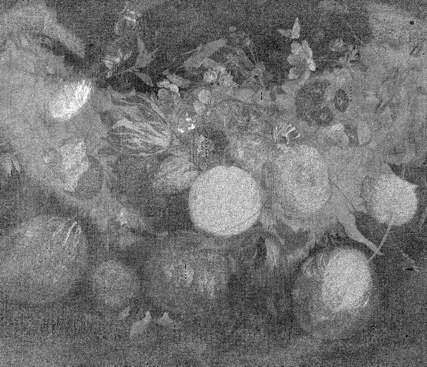}} &
        \multicolumn{2}{c}{\includegraphics[width = \imlen\textwidth]{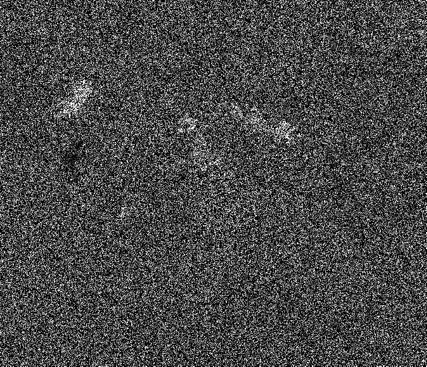}} &
        \multicolumn{2}{c}{\includegraphics[width = \imlen\textwidth]{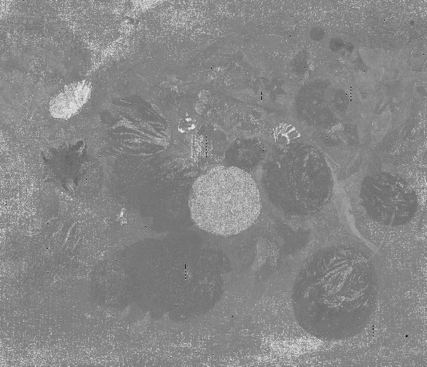}} &
        \multicolumn{2}{c}{\includegraphics[width = \imlen\textwidth]{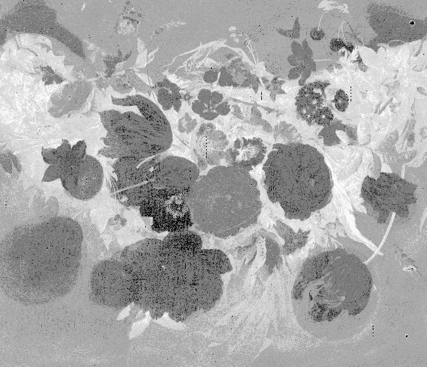}} \\
        
        \quad\; 420317 & (-11186) & \; 19160 & (-7326) & \quad 6178 & (-1250) & \quad 332 & (-183.83) & \quad 191172 & (110.99) & \quad 24.26 & (-14.91) & \quad \textbf{12.32} & (\textbf{-3.6893}) \\ \\
        
        \multicolumn{2}{c}{\rotatebox{90}{\small Ours} \includegraphics[width = \imlen\textwidth]{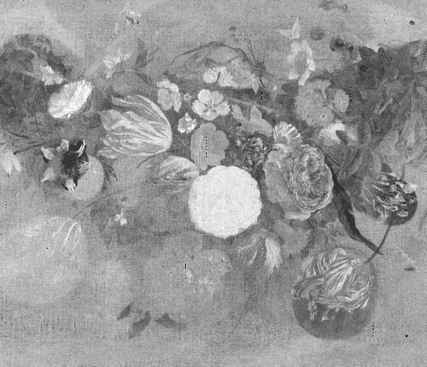}} &
        \multicolumn{2}{c}{\includegraphics[width = \imlen\textwidth]{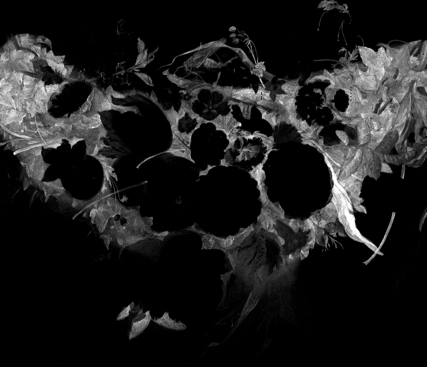}} &
        \multicolumn{2}{c}{\includegraphics[width = \imlen\textwidth]{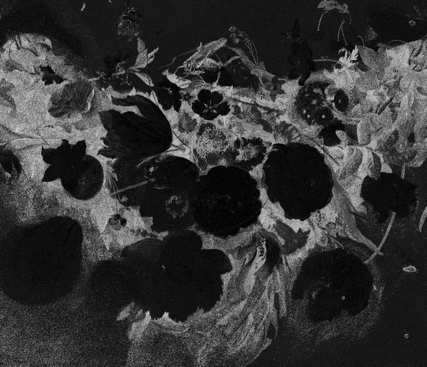}} &
        \multicolumn{2}{c}{\includegraphics[width = \imlen\textwidth]{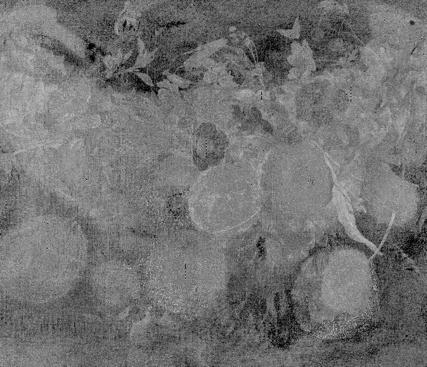}} &
        \multicolumn{2}{c}{\includegraphics[width = \imlen\textwidth]{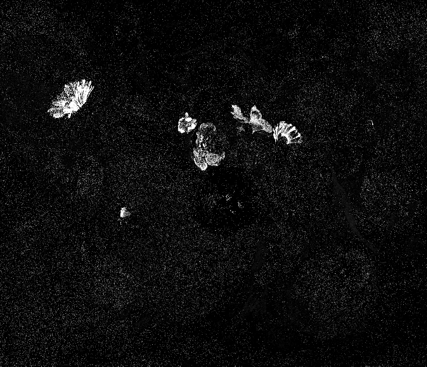}} &
        \multicolumn{2}{c}{\includegraphics[width = \imlen\textwidth]{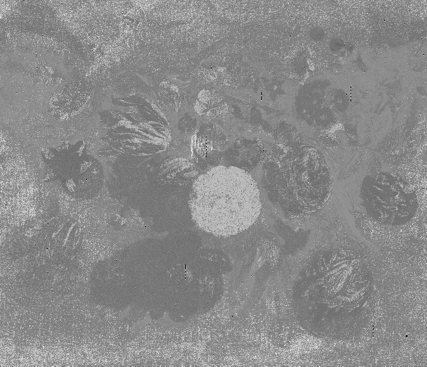}} &
        \multicolumn{2}{c}{\includegraphics[width = \imlen\textwidth]{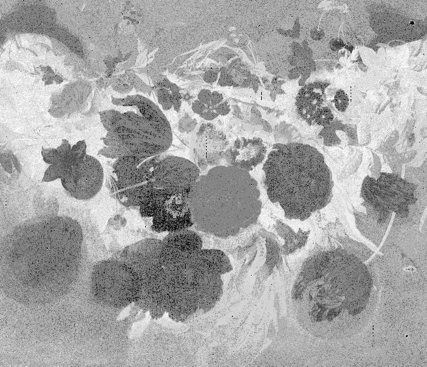}} \\
        
        \quad\; \textbf{21904} & (\textbf{-11373}) & \; \textbf{13747} & (\textbf{-7333}) & \quad 8240 & (-1247) & \quad \textbf{290} & (\textbf{-184.06}) & \quad 10792 & (-96.61) & \quad 23.49 & (-14.95) & \quad 12.42 & (-3.6813) \\ \\
        
        \multicolumn{2}{c}{\rotatebox{90}{\small Ground Truth} \includegraphics[width = \imlen\textwidth]{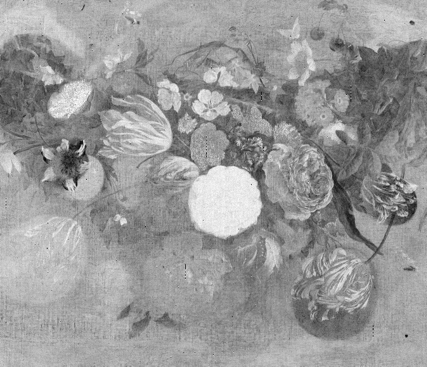}} &
        \multicolumn{2}{c}{\includegraphics[width = \imlen\textwidth]{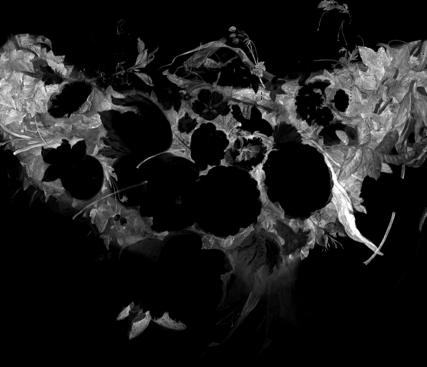}} & \multicolumn{2}{c}{\includegraphics[width = \imlen\textwidth]{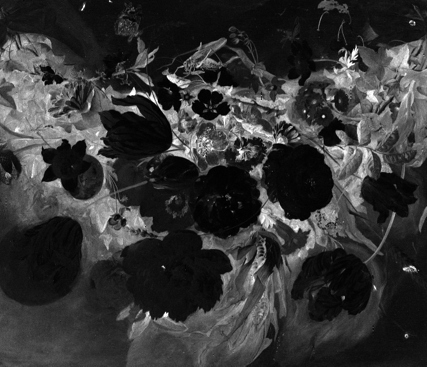}} & \multicolumn{2}{c}{\includegraphics[width = \imlen\textwidth]{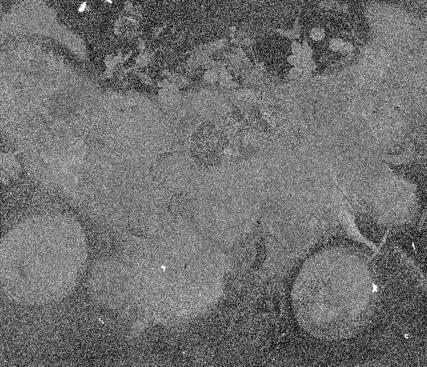}} & \multicolumn{2}{c}{\includegraphics[width = \imlen\textwidth]{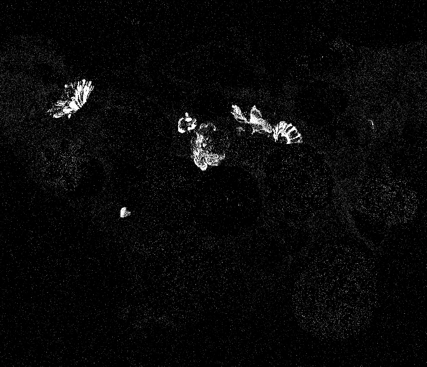}} & \multicolumn{2}{c}{\includegraphics[width = \imlen\textwidth]{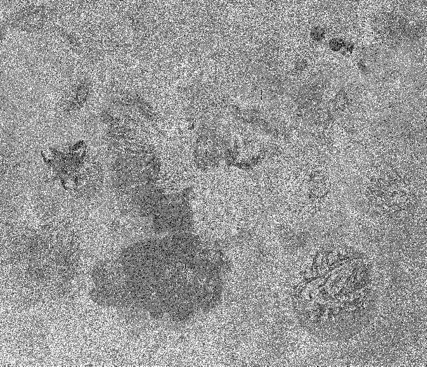}} & \multicolumn{2}{c}{\includegraphics[width = \imlen\textwidth]{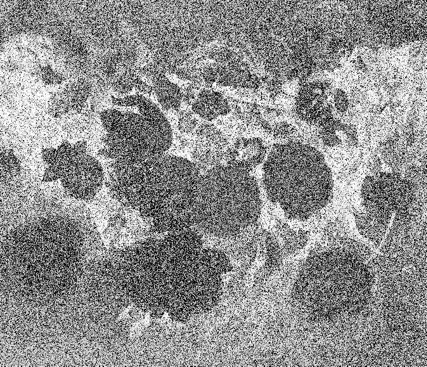}} \\
        
        \quad\; 0 & (-11380) & \; 0 & (-7342) & \quad 0 & (-1262) & \quad 0 & (-186.66) & \quad 0 & (-172.48) & \quad 0 & (-16.10) & \quad 0 & (-5.4176)
    \end{tabular}
    \caption{Visual comparison of seven elemental maps in the same display range per column. Row 1: the simulated fast XRF scan at $14.25$ ms/px. Row 2: PURE-LET denoising. Row 3: MCR-ALS denoising. Row 4: our denoising algorithm. Row 5: Ground truth XRF scan at $285$ ms/px. Numbers below are the PSNR and PNLL (in parentheses) respectively.}
    \label{fig:maps}
\end{figure*}
Recall that the XRF signal is a combination of elemental spectra.
Each element has its own unique XRF response that we can exploit for sparse coding, which has been shown to be effective in signal denoising~\cite{elad2010sparse}.
Let
\begin{equation}
    \label{eq:da}
    \bar{\xrf} = \dict \rep
\end{equation}
be a matrix reordering of $\xrf$ where $\dict \in \R^{\channels \times \atoms}_+$ is the dictionary with $\atoms$ non-negative atoms representing spectral responses, and $\rep \in \R^{\atoms \times \numel}_+$ is the sparse abundance matrix.
Each pixel of $\bar{\xrf} \in \R^{\channels \times \numel}_+$ is organized as $\numel = \height \cdot \width$ column vectors with $\channels$ features.
Learning the non-negative dictionary $\dict$ and abundance matrix $\rep$ provides both a spectrally smooth XRF volume and a more accurate representation of the chemical processes governing XRF data acquisition.

When scanning each pixel, photons of different energies arrive according to a Poisson sum model, which can be split into multiple independent Poisson processes~\cite{gallager2013stochastic}. 
Each process here describes the number of photon arrivals at each energy.
Each pixel we also assume to be spatially independent from one another.
Thus, we use the Poisson negative log likelihood (PNLL) loss as the data fidelity term:
\begin{equation}
    \label{eq:pnll_reg}
    \pnll \left( \xrf \right) = \sum_{h = 1}^\height \sum_{w = 1}^\width \sum_{c = 1}^\channels \xrf_{h, w, c} - \dwell \, \rate_{h,w,c} \cdot \ln \left( \xrf_{h, w, c} \right).
\end{equation}
Since $\rate$ is unknown, we instead try to best match the decomposition results with the data we record, namely:
\begin{equation}
    \label{eq:pnll}
    \pnll \left( \dict, \rep \right) = \sum_{c = 1}^\channels \sum_{n = 1}^{\numel} \left( \dict \rep \right)_{c, n} - \bar{\xrf}_{c, n} \cdot \ln \left( \left( \dict \rep \right)_{c, n} \right)
\end{equation}
which combines eq. (\ref{eq:da}) and eq. (\ref{eq:pnll_reg}) together.

The RGB image, $\rgb$, provides valuable and rudimentary insight into the spatial structure of each channel of the XRF volume.
Local areas similar in color likely have similar elemental profiles, and local areas of different colors likely have different elemental profiles.
The spatial gradient of $\rgb$ contains this information.
Define a total variation (TV) regularizer that adapts to the RGB gradient by:
\begin{align}
    \tv \left( \tilde{\rep} \right) = & \sum_{h = 1}^{\height - 1} \sum_{w = 1}^\width \sum_{c = 1}^\channels \frac{\rgbw^\height_{h,w}}{\dwell^2} \left( \tilde{\rep}_{h+1, w, c} - \tilde{\rep}_{h, w, c} \right)^2 + \nonumber \\
    \label{eq:tv}
    & \sum_{h = 1}^\height \sum_{w = 1}^{\width - 1} \sum_{c = 1}^\channels \frac{\rgbw^\width_{h,w}}{\dwell^2} \left( \tilde{\rep}_{h, w+1, c} - \tilde{\rep}_{h, w, c} \right)^2
\end{align}
where $\tilde{\rep} \in \R^{\height \times \width \times \atoms}$ is the volumetric representation of $\rep$ with the same pixel ordering as $\xrf$, and
\begin{align}
    \label{eq:tvh}
    \rgbw^\height_{h,w} &= \exp \left( - \beta \sum_{c = 1}^3 \left( \rgb_{h+1, w, c} - \rgb_{h, w, c} \right)^2 \right), \\
    \label{eq:tvw}
    \rgbw^\width_{h,w} &= \exp \left( - \beta \sum_{c = 1}^3 \left( \rgb_{h, w+1, c} - \rgb_{h, w, c} \right)^2 \right)
\end{align}
with $\beta > 0$ as a hyperparameter.
Adaptive weights $\rgbw^H$ and $\rgbw^W$ are large when the RGB gradient is small and vice-versa.
The division by $\dwell^2$ normalizes the TV regularizer result by time.
This TV regularizer is adapted from Dai \etal~\cite{dai2020adaptive} with the dwell time factored in.

We now define the full optimization problem as a weighted sum of eqs. (\ref{eq:pnll}) and (\ref{eq:tv}) with the addition of a weighted $l_0$ norm of $\rep$ to enforce sparsity constraints:
\begin{equation}
    \label{eq:opt}
    \dict^*, \rep^* = \argmin_{\dict, \rep \geq 0} \; \pnll \left( \dict, \rep \right) + \lambda_{\tv} \, \tv \left( \tilde{\rep} \right) + \lambda_{l_0} \, \lVert A \rVert_0.
\end{equation}
Once we have the optimized sparse representation of the signal, we can then find the optimized XRF photon rates by:
\begin{equation}
    \bar{\rate}^* = \frac{1}{\dwell} \, \dict^* \rep^*.
\end{equation}
$\bar{\rate}^*$ can be reshaped to the volumetric $\rate^*$ for analysis.

\subsection{Solution}
\label{sec:sol}

In order to solve eq.~(\ref{eq:opt}), we need a good initialization of $\dict$ and $\rep$ as well as a way to relax the $l_0$ norm.
We use K-means clustering of the spectral domain of all the pixels to initialize the dictionary.
A non-negative least squares fit of this dictionary and the XRF data quickly finds an optimal abundance matrix in a least-squares sense.
Elastic Net loss~\cite{zou2005regularization} and the Least Absolute Shrinkage and Selection Operator (LASSO)~\cite{tibshirani1996regression} replace the $l_0$ norm.
Elastic Net penalizes elements of $\rep$ by a combination of the $l_1$ and $l_2$ norms:
\begin{equation}
    \label{eq:elas}
    \loss_{\text{EN}} \left( \rep \right) = \alpha \, \lVert \rep / \dwell \rVert_2^2 + \left( 1 - \alpha \right) \lVert \rep / \dwell \rVert_1.
\end{equation}
LASSO sets values in $\rep$ below a threshold to zero and removes those elements from future updates.
The implemented optimization equation is then updated from eq.~(\ref{eq:opt}) as:
\begin{equation}
    \label{eq:opt2}
    \dict^*, \rep^* = \argmin_{\dict, \rep \geq 0} \; \pnll \left( \dict, \rep \right) + \lambda_{\tv} \, \tv \left( \tilde{\rep} \right) + \lambda_{\text{EN}} \, \loss_{\text{EN}} \left( \rep \right).
\end{equation}
Abundance matrix $\rep$ is updated using the Adam optimizer~\cite{kingma2014adam} until convergence which we define as when $i = i_{\text{min}} + j$ where $i$ is the current iteration number, $i_{\text{min}}$ is the iteration number with the smallest loss, and $j$ is a threshold.
Dictionary $\dict$ is updated also using Adam in an alternating fashion with $\rep$.
The optimal $\dict$ given a fixed $\rep$ cannot easily be found analytically due to the nature of the PNLL loss.
Thus, we turn to gradient descent-based methods.

\section{Experiments \& Results}
\label{sec:exp}

Jan Davidsz. de Heem's \textit{Bloemen en Insecten} as shown in fig.~(\ref{fig:rgb}) was scanned by de Keyser \etal~\cite{dekeyser2017jan}.
It consists of 2048 photon energy channels and has a resolution of $578 \times 673$ after registering the RGB image to the target XRF volume.
In our experiments, we treat this volume $\gt \in \R^{578 \times 673 \times 2048}$ as the ground truth photon count.
The dwell time per pixel for this acquisition was reported as $\dwell_\gt = 285$ ms/px.
Scanning an area of $578 \times 673$ spots with this dwell time would require over 30 hours of scanning.
The ground truth rate $\rate_\gt$ can be found by dividing $\gt$ by $\dwell_\gt$.

We identified 37 elements likely to compose a painting, leading to our choice of $\atoms = 37$ dictionary atoms.
Additionally, we set $\lambda_{\tv} = 10^{-2}$ and $\lambda_{EN} = 10^{-4}$.
Hyperparameters $\beta = 16$ of eqs.~(\ref{eq:tvh}) and~(\ref{eq:tvw}), and $\alpha = 0.2$ of eq.~(\ref{eq:elas}).

Using $\rate_\gt$, we simulate raster scans at various dwell times from a $5$-fold speedup ($57$ ms/px, about 6 hours scanning) to a $100$-fold speedup ($2.85$ ms/px, about 19 minutes scanning).
The MSE and mean PNLL error are our comparison metrics.
We compare against (1) PURE-LET2 with cycle-spinning ($5$ cyclic shifts) and $5$ Haar wavelet scales~\cite{luisier2010fast, coifman1995translation}, (2) our implementation of MCR-ALS~\cite{martins2016jackson} also with $37$ dictionary endmembers, and (3) the original simulated data without optimization.
PyMca~\cite{sole2007multiplatform}, a platform for XRF analysis, is used to generate the elemental maps for all the XRF volumes.

\begin{figure}[ht!]
    \centering
    \begin{subfigure}[b]{\linewidth}
        \includegraphics[width = \linewidth]{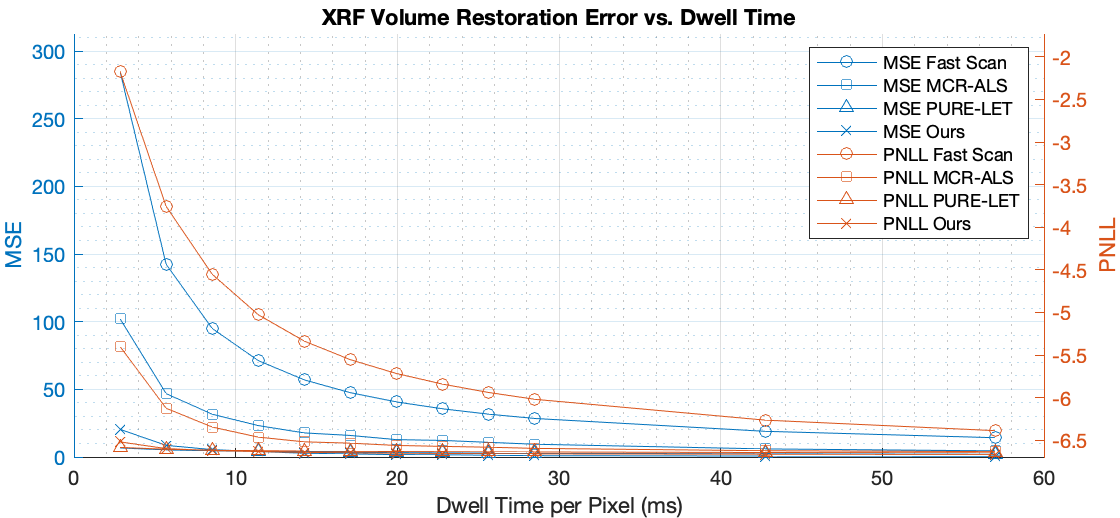}
        \caption{Errors of the optimized XRF volumes. The minimum PNLL for $\rate_\gt$ is $-6.6956$.}
        \label{fig:errall}
    \end{subfigure}
    \newline
    \newline
    \begin{subfigure}[b]{\linewidth}
        \includegraphics[width = \linewidth]{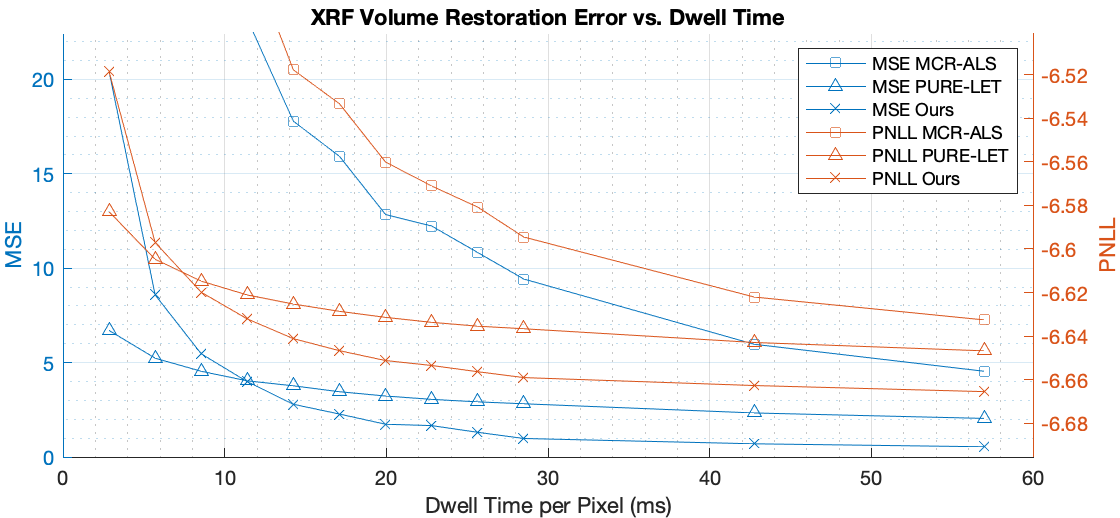}
        \caption{Close up error of the optimized XRF volumes only}
        \label{fig:erropt}
    \end{subfigure}
    \caption{Error of the XRF volumes}
    \label{fig:err}
\end{figure}

Fig.~(\ref{fig:maps}) shows some elemental maps of varying count rates and their corresponding MSEs and PNLLs.
The dwell time for those elemental maps is $0.05 \, \dwell_\gt \approx 14.25$ ms/px ($\approx 92$ minute scan time).
A volumetric comparison of the results are shown in the plots of fig.~(\ref{fig:err}).

We see in fig.~(\ref{fig:maps}) that our method performs best overall when considering error metrics and visual quality.
PURE-LET does well in reducing the error, especially for the lower dwell times as seen in fig.~(\ref{fig:erropt}), but qualitatively the maps tend to be oversmoothed (see Pb L3, Cl K, and Si K), making it difficult for further analysis.
MCR-ALS has the opposite problem in that the visual quality is high, but when examining fig.~(\ref{fig:errall}) and the elemental map metrics, the error on average is much higher than the other methods.
Noise tends to be most present in this method (see Pb L3 and As K), and some artifacts may be present.
For example, the Co K line, while visually appealing, is overestimated in some areas as opposed to something more uniform as the ground truth might suggest (see orange flower in the top left quadrant and central white flower).
Similarly in Cl K the same orange flower is overestimated as the ground truth shows a lower count rate at the bottom of the flower.

Our method generally shows well-denoised elemental maps both numerically and visually.
In terms of denoising the XRF volume as a whole, we outperform all the algorithms in MSE starting at about $11.5$ ms and in PNLL starting at about $7.5$ ms.
Visually, we are most consistent with the ground truth as well.

\section{Conclusion}
\label{sec:results}

We introduced a new method for denoising XRF volumes that combines a Poisson noise model with sparse dictionary learning.
Our algorithm outperforms methods designed for XRF denoising and Poisson denoising in general in quantitative and qualitative terms.
Speedups of a factor of 20 can not only ease time-related issues for accessing works of art, but could also open the opportunity for researchers to scan more paintings in a session.
Despite the speedup, our algorithm can still recover high-quality elemental maps more denoised than even the original data itself.
This allows more paintings to be analyzed for historical research and more quickly address conservation concerns.



\bibliographystyle{IEEEbib}

\end{document}